\documentclass[showpacs,secnumarabic,amsmath,amssymb,aps,twocolumn,superscriptaddress,prx]{revtex4-1}
\usepackage{mathtools}
\usepackage{array}
\usepackage{algorithm}
\usepackage{algpseudocode}
\usepackage{multirow}
\usepackage[pdftex]{graphicx} \graphicspath{{}}
\usepackage{grffile} 
\usepackage{comment}
\usepackage{makecell}
\usepackage{braket}
\usepackage{placeins}
\usepackage{diagbox}
\usepackage{bbding}
\usepackage{float}
\usepackage{afterpage}
\usepackage{amsthm}
\usepackage{graphicx}
\usepackage{graphics}
\usepackage{dcolumn}
\usepackage{bm}
\usepackage{tikz}
\usepackage{mathrsfs}
\usepackage{amssymb}
\usepackage{yhmath}
\usepackage[colorlinks,linkcolor=blue,urlcolor=blue,citecolor=blue]{hyperref}
\usepackage{tikz}

\usepackage{color}
\usepackage{xcolor}
\definecolor{ForestGreen}{RGB}{34, 139, 34}

\newcommand{\affA}{School of Physics, Peking University, 100871 Beijing, China.}
\newcommand{\affB}{Max Planck Institute for the Physics of Complex Systems, N\"othnitzer Str.~38, 01187 Dresden, Germany.}
\newcommand{\affC}{State Key Laboratory of Quantum Functional Materials, Department of Physics and Guangdong Basic Research Center of Excellence for Quantum Science, Southern University of Science and Technology, Shenzhen 518055, China}

\begin{document}
\title{{Engineering long-range and multi-body interactions via global kinetic constraints}}

\author{Runmin Wu}
 \affiliation{\affA}

 \author{Bing Yang}
 \affiliation{\affC}
 
\author{Pieter W. Claeys}
 \affiliation{\affB}

\author{Hongzheng Zhao}
\email{hzhao@pku.edu.cn}
 \affiliation{\affA}
	\date{\today}
    
\begin{abstract}
Long-range and multi-body interactions are crucial for quantum simulation and quantum computation. Yet, their practical realization using elementary pairwise interactions remains an outstanding challenge.
We propose an experimental scheme based on the Bose-Hubbard system with a periodic driving of the on-site energy and global-range density-density interactions, a setup readily implementable via cold atoms in optical lattices with cavity-mediated interactions. Optimally chosen driving parameters can induce global kinetic constraints, where tunneling rates are selectively suppressed depending on the particle number imbalance between all even and odd sites. This mechanism, together with the flexible tunability of local tunneling rates, provides efficient implementation schemes of a family of global controlled gates for quantum computation. We illustrate this scheme for the $N$-qubit Toffoli gate, circumventing the need for a two-body gate decomposition, and elaborate on the efficient preparation of entangled many-body states. 
\end{abstract}
	\maketitle
\let\oldaddcontentsline\addcontentsline
\renewcommand{\addcontentsline}[3]{}

\textit{Introduction.---}A central goal in quantum simulation is the realization of three-body and higher-order interactions beyond pairwise ones~\cite{will2010time,fedorov2012implementation,fletcher2017two,dai2017four,sala2020ergodicity,Petiziol2021QuantumSimulation,katz2022n,luo2024realization}.
One approach in their realization is Floquet engineering~\cite{aidelsburger2011experimental,struck2012tunable,greschner2014exploring,bukov2015universal,meinert2016floquet,eckardt2017colloquium,oka2019floquet}, where effective higher-order interactions can appear through the driving of native interactions~\cite{claeys2019floquet,choi2020robust,Petiziol2021QuantumSimulation}. 
This approach readily applies to a variety of present-day quantum simulator platforms~\cite{geier2021floquet,bluvstein2021controlling,scholl2022microwave,kalinowski2023non,sun2023engineering,jin2023fractionalized,martinez2016real,schweizer2019floquet,semeghini2021probing,satzinger2021realizing}, with applications in condensed matter~\cite{kitaev2006anyons} and high-energy physics~\cite{halimeh2025cold}.

A fundamental limitation in those protocols is that the engineered interactions generally remain spatially local and only act on a small number of lattice sites. Little is known on how to efficiently scale this up and engineer long-range, even global, higher-order interactions via Floquet driving. 
This question is especially pressing since universality of quantum computation and simulation requires the realization of multi-body controlled gates~\cite{shi_both_2003,aharonov_simple_2003,Nielsen_Chuang_2010}, which, however, present the main source of noise and decoherence in quantum devices. 

One famous example of a multi-body controlled gate is the Toffoli gate, which appears in fundamental quantum algorithms~\cite{grover_quantum_1997,roy_programmable_2020,haner_factoring_2017,gidney_how_2021} and quantum error correction~\cite{steane_error_1996,cory_experimental_1998,schindler_experimental_2011,reed_realization_2012,paetznick_universal_2013,yoder_universal_2016}. 
A single 3-body Toffoli gate requires at least five two-site controlled gates~\cite{barenco_elementary_1995,yu_five_2013,smith_leap_2023}, with this number growing at least linearly and generally quadratically for $N$-body Toffoli gates~\cite{shende_cnot-cost_2009}. This leads to significant interest in direct experimental realizations~\cite{monz_realization_2009,song_continuous-variable_2017,levine_parallel_2019,baekkegaard_realization_2019,baker_single_2022,kim_high-fidelity_2022,kim_high-fidelity_2022,glaser_controlled-controlled-phase_2023,dong_experimental_2024,nguyen_empowering_2024,liu_direct_2025} and highlights the immediate appeal of native realizations of multi-body controlled gates~\cite{wang_multibit_2001,rasmussen_single-step_2020,isenhower_multibit_2011,motzoi_linear_2017,khazali_fast_2020,bravyi_constant-cost_2022,nikolaeva_scalable_2024}. 
Here, we ask whether or not, and under which conditions, one can realize global-range and, importantly, multi-body interactions in a controllable and scalable way.

One way of circumventing the restrictions of spatial locality is through coupling to a non-local mode, e.g. a phonon in trapped ions or a photon in cavities, leading to noticeable progress in controlling long-range pairwise interactions~\cite{wang_multibit_2001,schutz2016dissipation,landig2016quantum,norcia2018cavity,rasmussen_single-step_2020,zhang2021observation,defenu2023long,ye2023universal,moon2024experimental,wu2024dissipative,guo2024site}.
However, modulating physical terms which do not commute with long-range interactions inevitably induces numerous analytically intractable higher-order processes and hence servere heating. It remains unclear how to select the desired multi-body interactions from many unwanted processes.

We overcome these difficulties by proposing a Floquet many-body system exhibiting effective {\it global} kinetic constraints (GKC). Local kinetic constraints have by now been identified in a range of strongly interacting systems~\cite{hudomal2020quantum,serbyn2021quantum,moudgalya2022quantum,chandran2023quantum,zhang2023many},
including Rydberg systems~\cite{bernien2017probing}, cold atoms in a large tilted potential~\cite{khemani2020localization,scherg2021observing,su2023observation,boesl2024deconfinement} and periodic modulation of pairwise interactions~\cite{zhao2020quantum,hudomal2020quantum}. The allowed dynamics are restricted depending on the local spin configuration, e.g., a spin is able to flip only if its neighbors are in a particular state, leading to the appearance of quantum many-body scars and fragmentation~\cite{serbyn2021quantum}.
While local kinetic constraints have been used to realize local controlled gates~\cite{isenhower_multibit_2011,levine_parallel_2019,jandura2022time,evered2023high}, generalizations to non-local higher-order interactions remain challenging. Also, the role of GKC in quantum dynamics is essentially unexplored.

In this work, we investigate systems exhibiting GKC, where the allowed dynamics depend on the global particle or spin configurations.
Concretely, we propose an cold-atom experiment in an optical lattice, with temporal modulation of the on-site energy and global-range density-density interaction. This model can be physically implemented using cavity-mediated global-range interactions~\cite{dogra2016phase,landig2016quantum,defenu2023long,chanda2024recent}. In the fast driving regime where heating can be well controlled~\cite{abanin2015exponentially,bukov2015prethermal,kuwahara2016floquet}, the system is described by an effective density-dependent tunneling processes, where the rate for a particle to tunnel depends on the particle number imbalance between all even and odd sites. These rates can be tuned through the drive, s.t. optimally chosen driving protocols can completely forbid certain transitions, realizing GKC.

For pedagogical reasons, we start from the 
2-qubit system and demonstrate that kinetically constrained tunneling can be used to implement a CNOT gate. Generalizing this setup to $N$ qubits, we illustrate a tower structure in the many-body Hilbert space, where transition rates can be classified via the total magnetization. 
Combined with the flexible tunability of local tunneling rates, this provides a versatile toolbox for the realization of non-local multi-body controlled gates. We illustrate it using a classical optimization method to identify an $N$-frequency driving profile that realizes an $N$-qubit Toffoli gate. Crucially, this scheme does not require a gate decomposition and therefore becomes particularly efficient for large $N$. We further present driving protocols for the preparation of globally entangled many-body states, including both the W state and GHZ state. Multi-body controlled gates here underpin the generation of multipartite entanglement, a valuable resource for quantum simulation and computation~\cite{friis_entanglement_2019}. We explicitly discuss possible experimental realizations before concluding.

\textit{The model.---}
We consider the extended Bose-Hubbard model (EBHM) on a chain
\begin{equation}
    \hat{H}(t){{=}}\sum_{\langle jk\rangle}\hat{c}_j^\dagger J_{jk}\hat{c}_k{+}F(t)\sum_{j}j\hat{n}_j{+}V(t)\sum_{k\neq j}\frac{(-1)^{k+j}}{2}\hat{n}_k\hat{n}_j,
\label{eq:bareHamiltonian}
\end{equation}
where $\hat{c}_j\,(\hat{c}_j^\dagger)$ annihilates (creates) a boson at site $j$, $\hat{n}_j{=}\hat{c}_j^\dagger\hat{c}_j$ is the occupation number operator, $J_{jk}$ is the bare tunneling rate between neighboring sites $\langle jk\rangle$, and we consider a tilted potential. The global-range interaction in the last term depends on the total imbalance between the number of bosons populating even and odd sites.
For simplicity we neglect the on-site contact interaction, but all proposed driving schemes can be directly extended to incorporate their effects. We consider a periodic drive $F(t)$ of the potential and of $V(t)$ in the interaction with zero time average, and exploit the freedom to choose suitable driving profiles to realize kinetically constrained models.  Eq.~\eqref{eq:bareHamiltonian} can be readily implemented
in cold-atom experiments~\cite{landig2016quantum}.

We consider strong driving where the driving amplitude can scale up with the driving frequency $\omega$. 

For concreteness, we consider the driving protocol 
$F(t)=3\omega F_3\cos(3\omega t){+}2\omega F_2\cos (2\omega t),\,V(t){=}\omega V_1\cos (\omega t)
$, where $\omega=2\pi/T$ with the driving period $T$. As detailed in End Matter, for a high-frequency drive, this protocol leads to the effective Hamiltonian~\cite{EndMatt}
\begin{eqnarray}
\hat{H}_{\mathrm{eff}}{=}\sum_{\langle jk\rangle}\hat{c}^{\dagger}_jJ_{jk}\,\mathcal{J}(F_3,F_2,V_1\hat{\Delta}_{jk}/2)\hat{c}_{k},
    \label{eq.effectiveHboson}
\end{eqnarray}
where we introduce the global density dependence
\begin{equation}
    \hat{\Delta}_{jk} 
    {=} 2(-1)^j\sum_{l}{(-1)^{l}}\hat{n}_l - \hat{n}_j+\hat{n}_k,
\label{eq.Aoperator}
\end{equation}
and define $\mathcal{J}(x,y,z)=\int_0^T\exp[i(x\sin 3\omega t+y\sin 2\omega t+z\sin \omega t]dt/T,$
a zeroth order 3-D Bessel function.

\begin{figure}[t]
    \centering
\includegraphics[width=\linewidth]{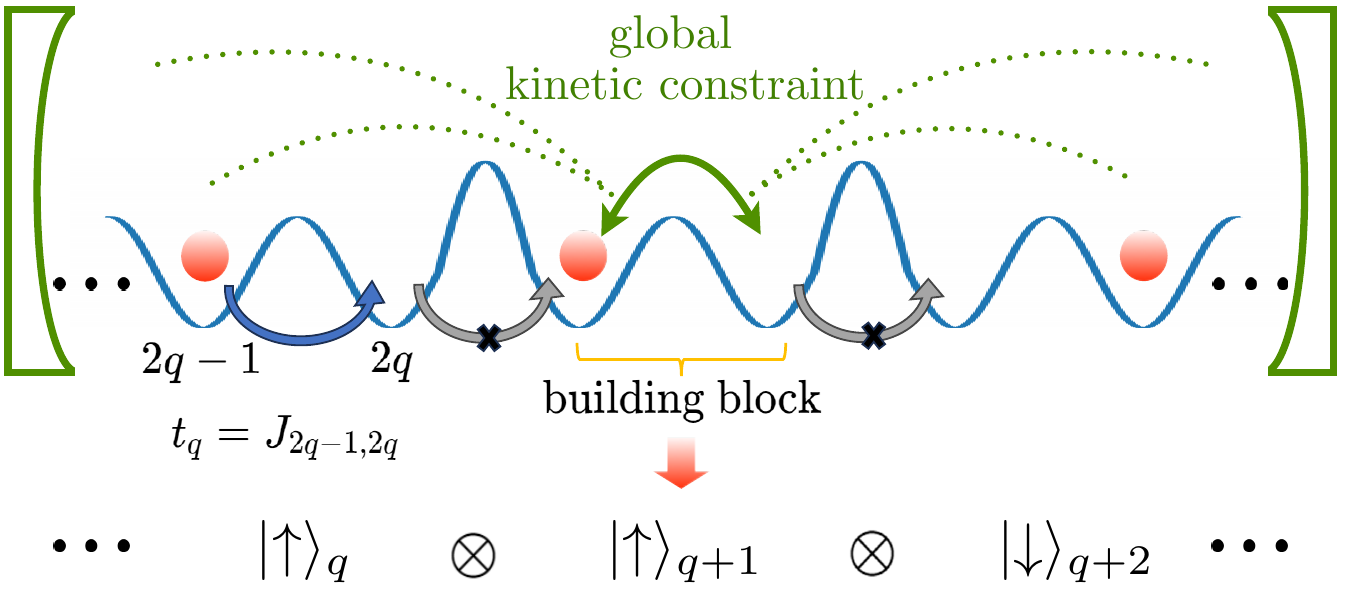}
    \caption{Illustration of the mapping from a bosonic system to a qubit system.
    One particle in two neighboring sites can be mapped to a single qubit. Particle hopping rate within a building block depends on the global particle distribution, realizing the global kinetic constraint (green); hopping between different blocks (grey) is forbidden. This can be physically implemented via driving cavity mediated interactions.} 
    \label{fig:1}
\end{figure}
The appearance of the density operators $\hat{\Delta}_{jk}$ in Eq.~\eqref{eq.effectiveHboson} indicates that the hopping rates now explicitly depend on the density distribution over the entire chain, since the operator $\sum_l(-1)^l \hat{n}_l$ quantifies the imbalance between even and odd sites. Importantly, although the original EBHM contains a global-range interaction, it is still a pairwise interaction. In contrast, through strong driving we obtain in Eq.~\eqref{eq.effectiveHboson} an effective multi-body interaction which simultaneously acts on all sites, fundamentally changing the locality of the underlying system. 

GKC can be realized by properly tuning the driving parameters to roots of Bessel functions, reminiscent of dynamical localization in Floquet systems~\cite{eckardt2017colloquium}. The key difference is that the Bessel function in Eq.~\eqref{eq.effectiveHboson} explicitly depends on the particle number distribution, allowing us to engineer desired state-dependent selection rules.

\textit{Pedagogical example: 2-qubit CNOT gate.---}
We now define a qubit subspace and rewrite the effective Hamiltonian [Eq.~\eqref{eq.effectiveHboson}] in the corresponding spin representation. We take the elementary building block to be two neighboring sites occupied by a single particle and identify
$|10\rangle_{\mathrm{BH}}{\to}|{\uparrow}\rangle$ and $|01\rangle_{\mathrm{BH}}{\to}|{\downarrow}\rangle.$
The particle number imbalance and the hopping within this building block reduce to the $\hat{\sigma}^z$ and $\hat{\sigma}^x$ operator respectively. One can enlarge the system size by concatenating $N$ different building blocks (see Fig.~\ref{fig:1}), where we introduce an infinite potential barrier to suppress the hopping between different blocks. 
Individual qubit subspaces remain well-defined and the spin representation consequently remains applicable.
For simplicity we first consider $N{{=}}2$ with
$
|1010\rangle_{\mathrm{BH}}{\to}|{\uparrow\uparrow}\rangle$, $|1001\rangle_{\mathrm{BH}}{\to}|{\uparrow\downarrow}\rangle$,
$|0101\rangle_{\mathrm{BH}}{\to}|{\downarrow\downarrow}\rangle$,
$|0110\rangle_{\mathrm{BH}}{\to}|{\downarrow\uparrow}\rangle$.
The particle-number imbalance between even and odd sites, appearing in the density-dependent operator $\hat{\Delta}_{jk}$, translates into the total magnetization $\sum_{l{=}1}^{2N}(-1)^{l}\hat{n}_l\rightarrow-\sum_{q{=}1}^N\hat{\sigma}_q^z,$ where we use $q$ to label the site index for qubits. 
The effective Hamiltonian further simplifies since we restrict each building block to be occupied by a single particle~\footnote{Here we used that, after applying the annihilation operator $\hat{c}_{k}$, the corresponding particle number operator within the same building block does not contribute to $\hat{\Delta}_{jk}$.}, resulting in
\begin{eqnarray}
    \hat{H}_{\mathrm{eff}}{{=}}t_1\mathcal{J}(F_3,F_2,-V_1\hat{\sigma}_2^z)\hat{\sigma}_1^x{+}t_2\mathcal{J}(F_3,F_2,-V_1\hat{\sigma}_1^z)\hat{\sigma}_2^x,
    \label{eq.2siteHamiltonian}
\end{eqnarray}
where we identify {$t_{q}\coloneqq J_{2q-1,2q}$}. 
Eq.~\eqref{eq.2siteHamiltonian} represents a two-body interaction of the $XZ$ form, with strength determined by both the bare hopping rates $t_1, t_2$ and the renormalization factor of the Bessel function.

\begin{figure}[t]
    \centering
\includegraphics[width=1\linewidth]{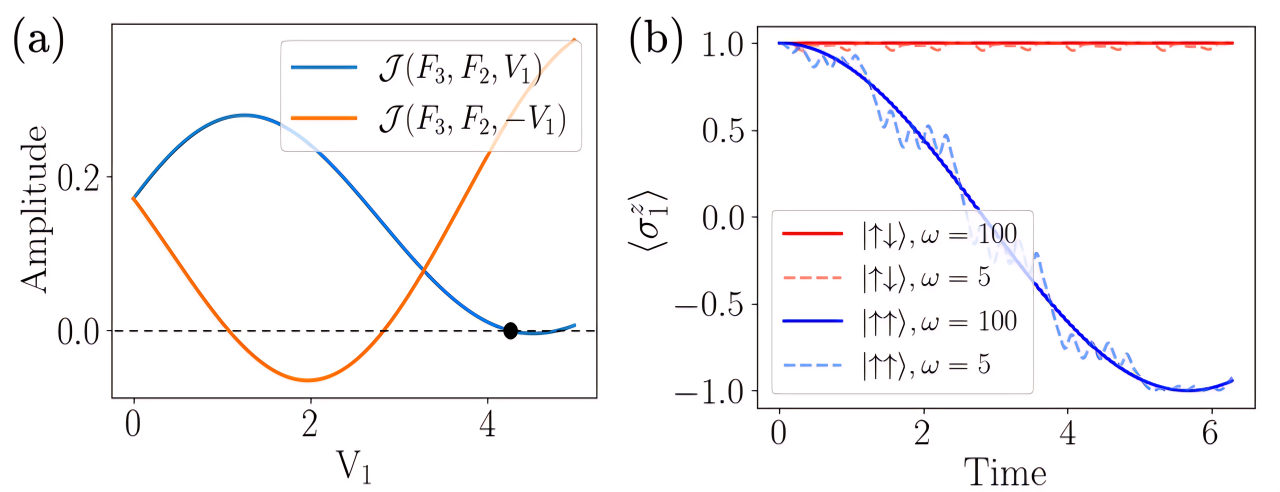}
    \caption{(a) Tuning transition rates by varying the driving amplitude $V_1$. One rate can be {completely} suppressed (leakage rate $10^{-8}$) while the other remains finite for $V_1\approx4.26$ (black dot) with $F_3{=}1, F_2{=}2.$ (b) Evolution of the expectation value of the magnetization of the first qubit. Only when the control qubit ($q{=}2$) is $|{\uparrow}\rangle$ can the target ($q{=}1$) evolve, realizing a CNOT gate. A low driving frequency (dashed lines) induces errors which become negligible in the high-frequency regime (solid lines). {We use $t_1{=}1$ and the driving parameters in panel (a) for the numerical simulation here.} }
    \label{fig:2}
\end{figure}
We treat $q{=}1$ and $q{=}2$ as the target and control qubit, respectively, and by setting $t_2{{=}}0$ we prevent transitions of the control qubit. The non-zero matrix elements read
\begin{equation}
\begin{aligned}
\hat{H}_{\mathrm{eff}}|{\uparrow\uparrow}\rangle & {=}\,t_1\mathcal{J}(F_3,F_2,-V_1)|{\downarrow\uparrow}\rangle,\\
    \hat{H}_{\mathrm{eff}}|{\uparrow\downarrow}\rangle & {=}\,t_1\mathcal{J}(F_3,F_2,V_1)|{\downarrow\downarrow}\rangle.
\end{aligned}
\label{eq.H_eff_twoqubit}
\end{equation}
By appropriately choosing the driving parameters, it is possible for targeted matrix elements to vanish, effectively `closing' hopping channels in the bosonic system and suppressing single-qubit transitions in the qubit system.
To construct a CNOT gate we hence require that
\begin{equation}
    \mathcal{J}(F_3,F_2,V_1){=}0,\quad \mathcal{J}(F_3,F_2,-V_1)\neq 0,
    \label{eq.conditiontwoqubit}
\end{equation}
s.t. the target qubit can flip only if the control qubit is in the state $|{\uparrow}\rangle$. 

We show the matrix elements as a function of $V_1$ in Fig.~\ref{fig:2}(a). The Bessel function has multiple roots, s.t. the conditions in Eq.~\eqref{eq.conditiontwoqubit} can be fulfilled. In practice, it is not necessary that these unwanted channels vanish exactly: a high-fidelity CNOT gate can be implemented in the presence of weak leaking channels, i.e. small but nonzero amplitudes of these rates. Fig.~\ref{fig:2} shows the numerical result of the expectation value of $\langle\hat \sigma_1^z\rangle$, starting from two different initial states. The initial state $|{\uparrow\uparrow}\rangle$ (blue) evolves while $|{\uparrow\downarrow}\rangle$ (red) remains stationary. The CNOT gate is implemented when $|{\uparrow\uparrow}\rangle$ evolves to $|{\downarrow\uparrow}\rangle$. While a finite driving frequency $\omega$ induces errors (dashed lines), these can be suppressed by increasing $\omega$, see details in End Matter. 

\begin{figure*}[t]
    \centering
\includegraphics[width=0.9\linewidth]{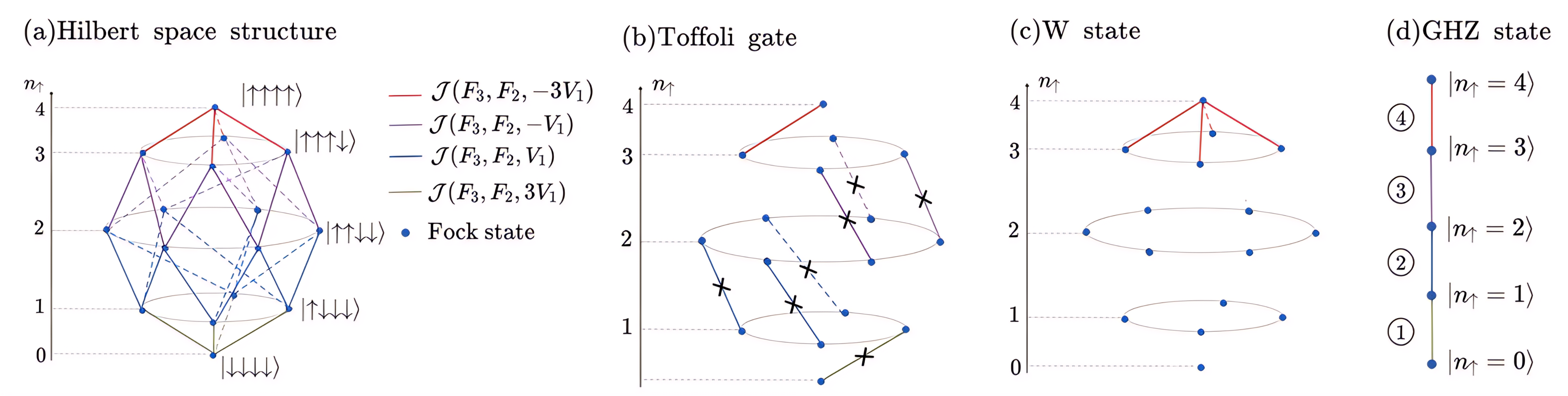}
    \caption{(a) Tower structure of the many-body Hilbert space. Fock states (solid dots) with the same ${n}_{\uparrow}$ live in the same layer.
    Only inter-layer transitions (colored lines) are non-vanishing, allowing us to divide the exponentially large Hilbert space into $N{+}1$ layers. The transitions are $N$-body interactions as their values rely on the total magnetization of a given state. (b) Hilbert space connections for the Toffoli gate, where all channels should be forbidden (black crosses), except the one connecting $|{\uparrow\uparrow\uparrow\uparrow}\rangle$ and $|{\downarrow\uparrow\uparrow\uparrow}\rangle$ (red line). (c) Hilbert space connections for the preparation of the W state. (d) Operational protocol for the preparation of the GHZ state. We use $N{=}4$ for illustration but the Hilbert space structure directly generalizes to large $N$.}
    \label{fig:hilbertspace}
\end{figure*}
\textit{Tower structure of the Hilbert space.---}
We generalize the discussion to $N$-qubit systems with a global $N$-body interaction, 
\begin{eqnarray}
\hat{H}_{\mathrm{\mathrm{eff}}}{=}\sum_{q{=}1}^Nt_q\mathcal{J}\Big(F_3,F_2,-V_1\sum_{p\neq q}\hat{\sigma}_p^z
\Big)\hat{\sigma}_{q}^x, \label{eq.effectiveHamiltonianNqubit}
\end{eqnarray}
where the total magnetization away from site $q$ now determines the spin-flip rate on site $q$.

The system features a simple tower structure in the many-body Hilbert space depending on the total number of spin-ups, ${n}_{\uparrow}$, of each Fock state. This is illustrated in Fig.~\ref{fig:hilbertspace} (a) using $N{{=}}4$ for concreteness, but the following discussion is generally applicable, and Fock states (solid dots) with the same ${n}_{\uparrow}$ live in the same layer (gray dotted circle). 
Colored lines denote the driving-induced renormalization of the bare hopping rate between two different Fock states. 
As $\hat{H}_{\mathrm{eff}}$ only contains a single-site flipping process, the matrix element is non-zero only if ${n}_{\uparrow}$ of two Fock states differ by 1. Only inter-layer transitions are hence non-vanishing, allowing us to divide the exponentially large Hilbert space into $N{+}1$ layers. The renormalization factors follow by noticing that the operator $\sum_{p\neq q}\hat{\sigma}_p^z$ in $\hat{H}_{\mathrm{eff}}$ evaluates to $2n_{\uparrow}{-}N{\pm}1$, where $+$(-) is taken if $n_{\uparrow}$ increases (decreases).
The renormalization factors between two layers are the same, as shown in Fig.~\ref{fig:hilbertspace} (a) by colored lines. 

\textit{$N$-qubit controlled gate.--}
Connections in the Hilbert space can now be made to vanish in two independent ways: either through the hopping process in the static Hamiltonian (setting $t_q{=}0$ for some $q$), resulting in cutting \emph{fixed} connections between \emph{all} layers, or by engineering the driving to close \emph{all} connections between two \emph{fixed} neighboring layers. We now exploit these two flexible methods to construct a $N$-qubit controlled gate. 

By setting $t_q{{=}}0$ for $q{\neq}1$ {in the static Hamiltonian}, we choose $q{{=}}1$ as the target qubit and the rest as control qubits. This leads to the disconnected Hilbert space as shown in Fig.~\ref{fig:hilbertspace} (b), where we again use $N{=}4$ for clarity. Note that we already erased all connections that flip the control qubits. Hence, in the absence of driving, all remaining transitions in Fig.~\ref{fig:hilbertspace} (b) correspond to a single-site flip of the target qubit.

A variety of $N$-qubit controlled gates can be engineered by completely suppressing some of the remaining channels. For illustration, we consider the $N$-qubit Toffoli gate where the target qubit can flip only if the control qubits are all spin up.
For this purpose, as shown in Fig.~\ref{fig:hilbertspace} (b), all channels should be forbidden (black crosses), except the one connecting the states $|{\uparrow\uparrow\uparrow\uparrow}\rangle$ and $|{\downarrow\uparrow\uparrow\uparrow}\rangle$, the red line. Therefore, the driving parameters should satisfy
\begin{equation}
    \mathcal{J}(F_3,F_2,-V_1){=}\mathcal{J}(F_3,F_2,V_1){=}\mathcal{J}(F_3,F_2,3V_1){=}0.
    \label{eq.conditionforToffoli}
\end{equation}
The task of finding suitable parameters can be converted to a classical optimization problem, where 
a cost function $g{=}|\mathcal{J}(F_3,F_2,-V_1)|+|\mathcal{J}(F_3,F_2,V_1)|+|\mathcal{J}(F_3,F_2,3V_1)|$ which quantifies the leakage channels, needs to be minimized. Through a gradient algorithm we achieve $g\approx 10^{-5}$ while $\mathcal{J}(F_3,F_2,-3V_1)$ remains finite~\cite{EndMatt}. We numerically verify this protocol in Supplementary Material (SM)~\cite{SupMatt} and achieve a high-fidelity realization of Toffoli gate in the high-frequency regime. 

This framework can be straightforwardly extended to construct a $N$-qubit Toffoli gate. For larger $N$, the complexity of the optimization task only scales linearly in $N$ as there are $N-1$ channels to be forbidden. We increase the dimension of the variable space {by including $N$ driving frequencies} and numerically minimize the cost function, which can be done efficiently~\cite{EndMatt}. It is worth highlighting the flexibility of the driving scheme: By selectively closing certain hopping channels, one can realize a family of $N$-qubit control gates beyond the standard Toffoli gate.

\textit{Preparation of entangled states.---}
Our protocol can efficiently prepare multipartite entangled many-body states including the W state and the GHZ state. We now set $t_q{=}1$ for all $q$ and use the effective Hamiltonian, Eq.~\eqref{eq.effectiveHamiltonianNqubit}. 
We again focus on the case with $N{=}4$ for a better elucidation of the Hilbert space structure, while generalizations are straightforward. The Hamiltonian now has permutation symmetry between arbitrary two sites, and thus we use $|{n_{\uparrow}}\rangle$ to represent the many-body state with a fixed value of $n_{\uparrow}$.

\begin{figure}[t]
     \centering    \includegraphics[width=1\linewidth]{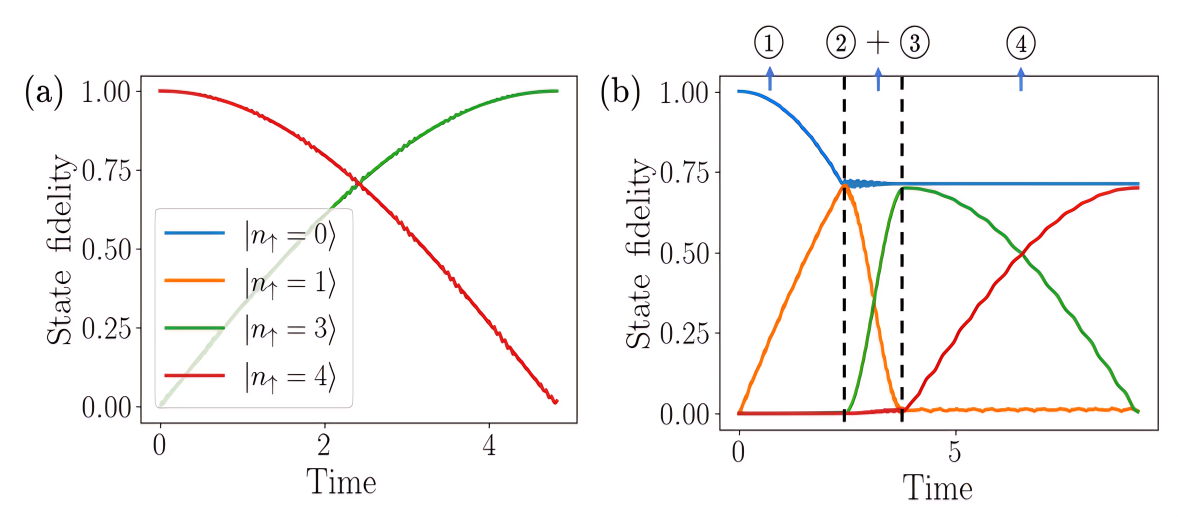}
     \caption{(a) Preparation of the W state. The overlap between the time evolved state and initial state (red) or the target W-state (green) is shown. We use $F_3{{=}}-6.38,F_2{{=}}-5.09,V_1{=}1.15$. (b) Preparation of the GHZ-state. We use $F_3=0,F_2=-1.01,V_1=0.82$ for steps 2 and 3. Overlaps with different states are plotted and the eventual GHZ state fidelity is $0.9996$. In both simulations $\omega=100$. }
\label{fig.numerics}
\end{figure}

For the W state, as shown in Fig.~\ref{fig:hilbertspace}, we require exactly the same set of conditions as for the realization of the Toffoli gate, Eq.~\eqref{eq.conditionforToffoli}, where only the red channels are non-vanishing. Suppose we prepare the initial state as $|{n_{\uparrow}{=}4}\rangle$, which sits on the top of the tower. As shown in Fig.~\ref{fig.numerics}(a), we evolve the system to $|{n_{\uparrow}{=}3}\rangle$, the target W-state, at the time $t{=}{\pi}/{4\mathcal{J}(F_3,F_2,-3V_1)}$. 

For the GHZ state, we consider an iterative protocol, shown in Fig.~\ref{fig:hilbertspace} (d), where we start from the initial state $|n_{\uparrow}{=}0\rangle$. First, we choose suitable driving parameters s.t. $\mathcal{J}(F_3,F_2,3V_1)$ (green) is the only non-zero channel. This leads to an effective two-level system, spanned by $|n_{\uparrow}{=}0\rangle$ and $|n_{\uparrow}{=}1\rangle$. 
After a suitable time, we realize the superposition $|n_{\uparrow}{=}0\rangle{+}|n_{\uparrow}{=}1\rangle$, effectively implementing the $\pi/2$ pulse.
We now close this green channel, s.t. the component $|n_{\uparrow}{=}0\rangle$ is isolated and remains unchanged. We turn on $\mathcal{J}(F_3,F_2,V_1)$ (blue) while closing all other channels, s.t. the $\pi$ pulse leads to the state $|n_{\uparrow}{{=}}0\rangle+|n_{\uparrow}{{=}}2\rangle$. Iterating this step leads to the target GHZ state, $|n_{\uparrow}{{=}}0\rangle{+}|n_{\uparrow}{{=}}4\rangle$. 
More efficient schemes can be developed by isolating an effective multi-level system during intermediate steps. Considering e.g. the subspace spanned by $|n_{\uparrow}{{=}}1\rangle,|n_{\uparrow}{{=}}2\rangle,|n_{\uparrow}{{=}}3\rangle$, we can find driving parameters that realize $\mathcal{J}(F_3,F_2,V_1){{=}}\mathcal{J}(F_3,F_2,-V_1)$ and suppress all other channels. As shown in Fig.~\ref{fig.numerics} (b), this degenerated three-level system can directly realize the transition from $|n_{\uparrow}{{=}}1\rangle$ to $|n_{\uparrow}{{=}}3\rangle$. 
This procedure directly applies to the $N-$qubit case, and one can iteratively prepare the GHZ state using $N/2$ steps for large $N$.

\textit{Experimental realization.---}
The Hamiltonian in Eq.~\ref{eq:bareHamiltonian} can be readily implemented in cold-atom experiments with the coupling to a high-finesse optical cavity~\cite{landig2016quantum}.
To prevent unwanted tunneling between neighboring building blocks, a bichromatic superlattice can isolate the system into individual double-well units~\cite{dai2017four}. 
The global-range interaction scales as $ {g^2}/{\Delta}$, where $g$ is the atom-cavity coupling strength, and $\Delta$ denotes the cavity detuning~\cite{landig2016quantum}.

In practice, this coupling can exceed $t_q$ by over two orders of magnitude, as required by our protocol. 
To realize the periodic drive, one may start with the laser red-detuned close to the cavity resonance, sweep the detuning toward $-\infty$ (eliminating the interaction), jump to a far blue detuning to invert the sign, and then sweep back to close the driving loop. In SM, we discuss concrete Hamiltonian parameters and show that our scheme lies within an experimentally feasible regime.
When combined with arbitrary single-qubit rotations~\cite{Impertro2024} and a Toffoli gate, this toolbox suffices for universal quantum computation~\cite{shi_both_2003,aharonov_simple_2003}.

The appearance of GKC is robust against static perturbations appearing in potential disorder and contact interactions. In SM, we also introduce a simpler protocol that achieves the same kinetic constraints solely by on-site potential modulation, significantly enhancing the experimental feasibility of our scheme. We can also decouple selected lattice sites from the cavity-induced interactions with other sites~\footnote{A tightly focused laser beam can be applied to the target lattice site to induce a large AC Stark shift, thereby detuning it from the optical transition. As a result, the detuning between the atomic transition and the pump light at this site becomes much larger than that of the other sites, thereby suppressing the cavity-induced interactions between a specific site and the rest of the lattice.}, allowing for the realization of a controlled gate on a flexible selection of lattice sites.

\textit{Discussion.---}
Various generalizations of the proposed protocol are possible. We extends this approach to qutrit systems by allowing a larger particle occupation in the elementary building block in~\cite{SupMatt}.
Also, although here we study 1D bosonic systems, generalization to higher-dimensional and fermionic systems is straightforward, where kinetic constraints may naturally lead to controlled fermionic gates
~\cite{konishi2021universal,zhang2021observation} relevant for fermionic simulation~\cite{kivlichan_quantum_2018,google_ai_quantum_demonstrating_2020}.
Long-range interactions have been used to optimize spatial searches motivated by Grover's algorithm~\cite{king_optimal_2025}, and it would be interesting to explore possible speed-ups through the use of multi-body interactions. Another natural follow-up would be to use these interactions to realize $N \geq 3$-qubit quantum error correcting codes~\cite{steane_error_1996,cory_experimental_1998,schindler_experimental_2011,reed_realization_2012,paetznick_universal_2013,yoder_universal_2016}.
Our work thus paves the way for future application of quantum simulation and computation, whenever multi-body interaction is involved. 

Kinetic constraints can break the ergodicity of the quantum many-body evolution via fragmentation~\cite{sala2020ergodicity}. So far, the exploration of constrained dynamics is mostly restricted to local constraints. Our work motivates a systematic study of ergodicity-breaking and transport phenomenon, when a global constraint is present~\cite{nicolau2025fragmentationzeromodescollective}. Finally, the interplay between the emergent Dicke symmetry in systems with global interactions and kinetic constraints is also of particular interest~\cite{defenu2023long}.

\textit{Acknowledgments.---}
We thank Wenlan Chen, Xiangliang Li for insightful discussions on the possible experimental realization.
This work is supported by {Quantum Science and Technology-National Science and Technology Major Project
(Grant No. 2024ZD0301800),} and by the National Natural Science
Foundation of China (Grant No. 12474214), and by ``The Fundamental Research Funds for the Central Universities, Peking University”, and by ``High-performance Computing Platform of Peking University".
P.W.C. acknowledges support from the Max Planck Society.
B. Y. acknowledges support from the NSFC (Grant No. 12274199), Shenzhen Science and Technology Program (Grant No. KQTD20240729102026004), Guangdong Major Project of  Basic and Applied Basic Research (Grant No. 2023B0303000011) and Guangdong Provincial Quantum Science Strategic Initiative (Grant No. GDZX2304006, Grant No. GDZX2405006).

\clearpage
\twocolumngrid
\part*{\scalebox{0.6}{End Matter}}
\addcontentsline{toc}{part}{End Matter}

\section{Derivation of the Hamiltonian in the rotating frame}
Here we present the details in the derivation of the effective Hamiltonian in a rotating frame.
We consider the unitary transformation $$
U(t){=}\exp\left[i\theta(t)\sum_j\epsilon_j\hat{n}_j+{i\beta(t)}\sum_{j\neq k}{(-1)^{j+k}}\hat{n}_j\hat{n}_k/2\right],$$
where $\theta(t)=\int_0^t d\tau F(\tau),\, \beta(t)=\int_0^{t}d\tau V(\tau)$.
The Hamiltonian in the rotating frame can be obtained as $$\widetilde{H}{=}U(t)\hat{H}U^\dagger(t)- iU(t)\frac{\partial U^\dagger(t)}{\partial t}.$$ This leads to
        $\widetilde{H}(t) {=}\sum_{\langle jk\rangle}\hat{c}_j^\dagger\hat{A}_{jk}(t)\hat{c}_k,$
where the global density dependence is contained in the operator
\begin{align}
        \hat{A}_{jk}(t)&{=} J_{jk}\exp[i\theta(t)(\epsilon_j-\epsilon_k)]\exp\left[i\beta(t)\hat{\Delta}_{jk}/2\right],
\end{align}
where $\epsilon_j=j$, $\hat{\Delta}_{jk} {=} 2(-1)^j\sum_{l}{(-1)^{l}}\hat{n}_l - \hat{n}_j+\hat{n}_k,$
and we used $|j-k|=1$.
The Hamiltonian
$\tilde{H} (t)$ in this frame admits a perturbative high-frequency expansion even for strong driving~\cite{bukov2015universal}. The lowest order contribution to the high-frequency expansion reads
$\hat{H}_{\mathrm{eff}}{=}\int_{0}^{T}\tilde{H}(t)dt/T,$ with $T$ the driving period. Higher order contributions are negligible in the high-frequency regime, $\omega{\gg} J_{jk}$, with $\omega=2\pi/T$.

For the driving protocol considered in the main text, 
$F(t)=3\omega F_3\cos(3\omega t)+2\omega F_2\cos (2\omega t),\,V(t)=\omega V_1\cos (\omega t),
$
leading to time-dependent phase factors $\theta(t){=}F_3\sin (3\omega t)+ F_2\sin (2\omega t),\,\beta(t){=}V_1\sin (\omega t)$ in Eq.~\eqref{eq.Aoperator}. 
The lowest-order contribution to the effective Hamiltonian corresponds follows as
\begin{eqnarray}
\hat{H}_{\mathrm{eff}}{=}\sum_{\langle jk\rangle}\hat{c}^{\dagger}_jJ_{jk}\mathcal{J}^{3\omega,2\omega,\omega}(F_3,F_2,V_1\hat{\Delta}_{jk}/2)\hat{c}_{k},
\end{eqnarray}
returning Eq.~\eqref{eq.effectiveHboson} from the main text.

\section{Classical optimization and generalization to $N-$qubit gate}
We provide details for the optimization algorithm used to identify the optimal driving parameters for the controlled gate. As discussed in the main text, the driving parameters for the 4-qubit Toffoli gate should satisfy Eq.~\eqref{eq.conditionforToffoli}. Solving this equation can be transformed into an optimization problem by minimizing the cost function
\begin{equation}
\nonumber
g{=}|\mathcal{J}(F_3,F_2,-V_1)|+|\mathcal{J}(F_3,F_2,V_1)|+|\mathcal{J}(F_3,F_2,3V_1)|,
\end{equation}
which quantifies the leakage channels.
We employ the Conjugate Gradient Method (CG) for minimization. Upon specifying an initial condition, CG is capable of locating a local minimum of $g$ within the vicinity of this starting point. By scanning over different initial conditions, multiple local minima in the broader parameter spaces can be obtained, and we use the optimal solution obtained after a few trials.

We now extend this framework to the $N$-qubit  Toffoli gate. For larger $N$, we increase the dimension of the variable space to ensure the existence of roots of Bessel functions. The complexity of this optimization task only scales linearly in $N$ as there are $N-1$ channels to be forbidden. 
Concretely, we consider the $N$-frequency drive
$F(t)=N\omega F_N\cos (N\omega t)+\dots+ 3\omega F_{3}\cos(3\omega t)+2\omega F_{2}\cos (2\omega t),V(t) =\omega V_1\cos (\omega t)$, 
leading to an effective Hamiltonian expressed in terms of $N$-dimensional Bessel functions
\begin{equation*}
H_{\mathrm{eff}}{{=}}\sum_{q{=}1}^N t_q\mathcal{J}^{N}(F_N,\dots,F_{2},-V_1\sum_{p\neq q}\hat{\sigma}_p^z)\sigma_q^x,
\end{equation*}
where the superscript $N$ denotes the dimension of the Bessel function, defined as 
\begin{equation}
\mathcal{J}^N(F_N, \dots F_2,F_1)=\frac{1}{T}\int_0^T\exp\left[i \sum_{j=1}^N F_j \sin(j \omega t)\right]dt.
\end{equation}
We also require $t_q=0$ for $q\neq 1$ to fix $q=1$ as the target qubit. This Hamiltonian can be used to achieve the $(N{+}1)-$qubit Toffoli gate with sufficiently small leaking channels. Now we use the cost function
\begin{equation}
g{=}\sum_{n{=}1}^{N}|\mathcal{J}^N(F_N,F_{N-1},...,F_2,(2n-N)V_1)|,
\end{equation}
which can be efficiently minimized using the CG algorithm. In Table.~\ref{Tab:1} we present the optimized driving parameters for at most 9 qubits, where $\bold{F}$ defines the driving profile
\begin{equation}
    \bold{F}{=}(F_{N},F_{N-1},...,F_2,V_1).
\end{equation}
The complexity of extending this optimization task to larger system sizes scales linearly in $N$, s.t. the optimal driving profile can be efficiently identified numerically. Also, this cost function (leakage rate) can be systematically suppressed by improving the classical optimization outcomes, such as by using better algorithms or increasing the optimization duration. Importantly, for a specific control gate and a fixed number $N$, this classical optimization process only needs to be implemented once. Once the optimized results have been identified, the same driving profile can always be used in the experiment without incurring additional quantum resource costs.

\section{Error analysis}
Experimental implementations of driven quantum systems are fundamentally constrained by the achievable drive frequency $\omega$. This limitation induces errors between the actual Floquet dynamics and the evolution induced by the target effective Hamiltonian $\hat{H}_{\mathrm{eff}}$. Here we show that this deviation can be parametrically suppressed as $\mathcal{O}(\omega^{-1})$ by using a larger driving frequency.

\begin{figure*}[h]
    \centering
\includegraphics[width=0.9\linewidth]{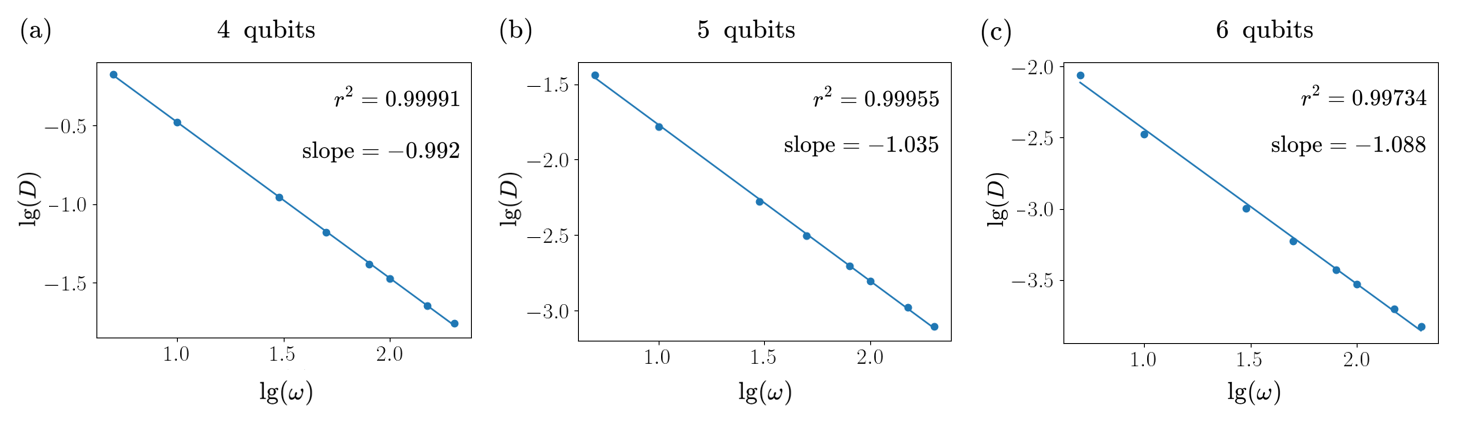}
    \caption{Linear fit of $\lg(\omega)$ vs. $\lg(D)$. We observe a linear dependence with slope $-1$ between $\lg(\omega)$ and $\lg(D)$ for 4-, 5-, and 6-qubit case. $r^2$ denotes the correlation coefficient in this linear fitting. 
    }
    \label{fig:S2}
\end{figure*}

To quantify this error, we consider the time-averaged deviation
\begin{equation}
    D=\frac{1}{t_{\mathrm{max}}}\int_0^{t_{\mathrm{max}}}dt\,|\langle\sigma_0^z\rangle_{\mathrm{real}}(t)-\langle\sigma_0^z\rangle_{\mathrm{eff}}(t)|,
\end{equation}
where $\langle\sigma_0^z\rangle_{\mathrm{real}}(t)$ is the expectation value of $\sigma_0^z$ at time $t$ and the subscript `$\mathrm{real}$' indicates that the time evolution is governed by the Floquet drive. Similarly, $\langle\sigma_0^z\rangle_{\mathrm{eff}}(t)$ denotes the results obtained by the effective Hamiltonian. We start from the initial state $|{\uparrow\uparrow...\uparrow\rangle}$ and consider the driving protocol for the Toffoli gate where we fix the evolution time $t_{\mathrm{max}}=5$. In Fig.~\ref{fig:S2} we plot the numerical results for different driving frequencies $\omega$ and different system sizes. We use a double log scale and identity a power-law scaling dependence between the error and the frequency, with the scaling exponent being close to 1. This implies that the first-order perturbation in the Floquet-Magnus expansion dominates the error, which can be linearly suppressed using larger $\omega$.

\onecolumngrid

\begin{table*}[t]
    \centering
   \begin{tabular}{|c|c|c|c|}
\hline
   Number of qubits $N+1$ & Driving profile $\bold{F}{=}(F_{N},F_{N-1},...,F_2,V_1)$ & Cost function $g$ \\

   \hline
   4&(-11.638,1.770,-3.065)&$10^{-8}$\\
   \hline
   5&(-6.620,-9.154,4.990,2.103)&$10^{-8}$\\
   \hline
   6&(-9.317,-4.385,-7.103,-5.065,-0.301)&$2\times10^{-3}$\\
   \hline
   7&(-7.074,-7.197,-6.605,-7.058,-6.822,-0.107)&$3\times 10^{-3}$\\
   \hline
   8&(-8.192,-7.452,-7.003,-7.074,-7.465,-6.280,-0.220)&$3\times 10^{-3}$\\
   \hline
   9&(-6.649,-6.262,-5.360,-7.220,-7.341,-6.338,-5.904,-0.437)&$5\times 10^{-3}$\\
   \hline
\end{tabular}
    \caption{Driving profile for the realization of the $(N+1)$-qubit Toffoli gate. The minimization of the leakage channel through the cost function $g$ can be efficiently performed numerically.}
    \label{Tab:1}
\end{table*}

 \let\addcontentsline\oldaddcontentsline
	\cleardoublepage
	\onecolumngrid
 \begin{center}
\textbf{\large{\textit{Supplementary Material} \\ \smallskip
	for ``Engineering long-range and multi-body interactions via global kinetic constraints" }}\\
		\hfill \break
		\smallskip
	\end{center}
	
	\renewcommand{\thefigure}{S\arabic{figure}}
	\setcounter{figure}{0}
    \renewcommand{\thesection}{SM\;\arabic{section}}
	\setcounter{section}{0}
	\tableofcontents
 \setcounter{page}{1}

 \section{Numerical verification of the Toffoli gate}
\begin{figure}[h]
    \centering
    \includegraphics[width=0.4\linewidth]{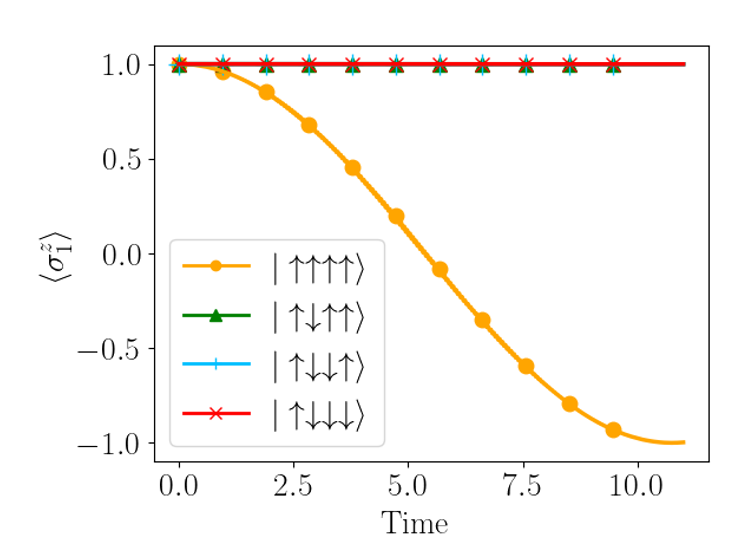}
    \caption{Time evolution of the expectation value of the magnetization of the first qubit $\langle\hat\sigma_1^z\rangle$, starting from 4 different initial states. We choose drive frequency $\omega=100$ and driving profile $F_3=-6.38,F_2=-5.09,V_1=1.15$. Only when the control qubits are all spin up can the target qubit evolve, demonstrating that the driving protocol implements the targeted Toffoli gate. }
    \label{fig:5}
\end{figure}

We numerically verify the protocol for the realization of the Toffoli gate by simulating the time-dependent Bose-Hubbard system. We consider 4 different initial states and use a high-frequency drive. The evolution of $\langle\sigma_1^z\rangle$ is plotted in Fig.~\ref{fig:5}. When all control qubits are spin up (yellow), the target qubit evolves. In contrast, for other initial states, the system remains unchanged, indicating the successful implementation of the Toffoli gate.

\section{Controlled gate for qutrit systems}
In addition to the quantum gates in qubit systems, the protocol can be generalized to qutrit gates. In this section, we first clarify the mapping from the bosonic system to the qutrit system, then we provide a protocol to realize a two-qutrit CNOT gate. Finally, we illustrate the possibility of controlling a single qutrit via multiple qubits.

A single qutrit can be defined by using a double well occupied by two bosonic atoms, returning an effective 3-level system as
\begin{equation}
|20\rangle_{\mathrm{BH}}\rightarrow|{2}\rangle_{\mathrm{Q}},\quad |11\rangle_{\mathrm{BH}}\rightarrow|{1}\rangle_{\mathrm{Q}}, \quad |02\rangle_{\mathrm{BH}}\rightarrow|{0}\rangle_{\mathrm{Q}},
\end{equation}
where we use the subscript Q to denote the qutrit basis.
Similar to the qubit system discussed in the main text, we can concatenate $N$ elementary building blocks to enlarge the system size. For simplicity we first consider $N=2$, where we have the mapping between the basis states $|2020\rangle_{\mathrm{BH}}\rightarrow|22\rangle_{\mathrm{Q}}$, $|2011\rangle_{\mathrm{BH}}\rightarrow|21\rangle_{\mathrm{Q}}$,$|2020\rangle_{\mathrm{BH}}\rightarrow|22\rangle_{\mathrm{Q}}$, etc. The discussion is similar to the qubit system, but there are two main differences. First, the particle occupation within two neighboring sites $j,k$ now contributes to the operator $\hat{\Delta}_{jk}$. This is not the case in qubit systems where the operator $\hat{c}_k$ annihilates the single particle in the double well. Hence, the state of the qutrit will also influence its own flipping rate. Second, there are 3 different internal states for a qutrit, and hence the number of hopping channels to be controlled is larger than the qubit case. In fact, for the $N-$qutrit system, there are $4N-2$ hopping channels to be considered in the optimization procedure if one wants to realize the multi-body controlled gate, which is however still linear in system size. 

\par
Similar to the discussion in the main text, we use $q$ to label the site index of the qutrit, and use $t_q$ as the bare hopping rate within the double well corresponding to site $q$. We treat $q=1$ and $q=2$ as the target and control qutrit, respectively, by setting $t_2=0$. We consider the `qutrit CNOT gate', where the target qutrit can only flip when the control qutrit is in the state $|2\rangle$. For this purpose, we consider a 4-frequency driving with tunable amplitudes. There are 6 possible non-zero hopping channels,
    \begin{align}
        & \hat{H}_{\mathrm{eff}}|22\rangle_{\mathrm{Q}}=\sqrt{2}t_1\mathcal{J}(F_4,F_3,F_2,-\frac{5}{2}V_1)|12\rangle_{\mathrm{Q}},\\
       &\hat{H}_{\mathrm{eff}}|20\rangle_{\mathrm{Q}}=\sqrt{2}t_1\mathcal{J}(F_4,F_3,F_2,\frac{3}{2}V_1)|10\rangle_{\mathrm{Q}},\\
      &\hat{H}_{\mathrm{eff}}|21\rangle_{\mathrm{Q}}=\sqrt{2}t_1\mathcal{J}(F_4,F_3,F_2,-\frac{1}{2}V_1)|11\rangle_{\mathrm{Q}},\\
      &\hat{H}_{\mathrm{eff}}|00\rangle_{\mathrm{Q}}=\sqrt{2}t_1\mathcal{J}(F_4,F_3,F_2,\frac{5}{2}V_1)|10\rangle_{\mathrm{Q}},\\
&\hat{H}_{\mathrm{eff}}|02\rangle_{\mathrm{Q}}=\sqrt{2}t_1\mathcal{J}(F_4,F_3,F_2,-\frac{3}{2}V_1)|12\rangle_{\mathrm{Q}},\\
&\hat{H}_{\mathrm{eff}}|01\rangle_{\mathrm{Q}}=\sqrt{2}t_1\mathcal{J}(F_4,F_3,F_2,\frac{1}{2}V_1)|11\rangle_{\mathrm{Q}}.
\end{align}
To construct a qutrit CNOT gate, we require 4 hopping channels to be suppressed:
\begin{equation}
    \mathcal{J}(F_4,F_3,F_2,\frac{3}{2}V_1)=\mathcal{J}(F_4,F_3,F_2,-\frac{1}{2}V_1)=\mathcal{J}(F_4,F_3,F_2,\frac{5}{2}V_1)=\mathcal{J}(F_4,F_3,F_2,\frac{1}{2}V_1)=0.
\end{equation}
Numerically, we can again define a cost function and minimize this to obtain a suitable drive profile as:
\begin{equation}
    F_4=-7.624,F_3=-7.092,F_2=0.592,V_1=-6.403.
\end{equation}
\par
We finally consider an alternative setup where we use mutiple qubits to control a single qutrit, again focusing on the Toffoli gate for concreteness: only when all control qubits are spin up can the qutrit flip. Since the state of the target qutrit contribute to its hopping intensity, there are $2(N+1)$ hopping channels for $N$ control qubits, of which we need to close $2N$. We hence consider a driving protocol with $2N$ drive frequencies to ensure the existence of a solution. Numerically, we again minimize the cost function for $N$ control qubits 
\begin{equation}
    g{=}\sum_{n{=}1}^{2N}\left|\mathcal{J}^{2N}\left(F_{2N},F_{2N-1},...,F_2,({2N+3}/{2}-n)V_1\right)\right|.
\end{equation}
We present the numerical result for $N$ control qubits in Table \ref{Tab:2}, with $N$ up to 4. We still use a vector $\bold{F}$ to represent the drive profile,
\begin{equation}
    \mathbf{F}{=}(F_{2N},F_{2N-1},...,F_2,V_1).
\end{equation}
We observe that we can again efficiently obtain driving protocols for which these channels can be systematically closed.

\begin{table}[!h]
    \centering
   \begin{tabular}{|c|c|c|}
   \hline
  \toprule Number of controlled qubits & Driving profile $\mathbf{F}$ & Cost function $g$ \\
  \hline
   2&(-7.193,-7.033,-6.811,4.794)&$10^{-8}$ \\
   \hline
   3&(-7.480,-7.134,-6.619,-6.825-6.760,0.403)&$2\times 10^{-4}$\\
   \hline
   4&(-7.104,-6.909,-6.046,-7.689,-6.767,-6.1423,-4.383,-1.026)&$3\times 10^{-4}$ \\
   \hline
\end{tabular}
    \caption{Driving profile for qutrit controlled gates via $N$ control qubits, with $N$ up to 4.  The value of the cost function $g$ quantifies the leakage channels.}
    \label{Tab:2}
\end{table}

\section{Alternative Floquet protocol}

Directly driving long-range interactions can be experimentally challenging and
resource-intensive (but not impossible). Here, we introduce a simpler protocol that achieves the same global-density–dependent tunneling solely by modulating the site potential. For concreteness, we consider the Hamiltonian
\begin{equation}
     \hat{H}(t){{=}}\sum_{\langle jk\rangle}\hat{c}_j^\dagger J_{jk}\hat{c}_k{+}F(t)\sum_{j}j\hat{n}_j{+}\sum_{k\neq j}\frac{V_1(-1)^{k+j}}{2}\hat{n}_k\hat{n}_j,
\end{equation}
where the interaction strength $V_1$ is now static and the amplitude of site potential is modulated $F(t)=F_1\,\omega\cos\omega t+F_2\,2\omega\cos2\omega t+F_3\,3\omega\cos3\omega t$. We consider the unitary transformation
\begin{equation}
    U(t){=}\exp\left[i\theta(t)\sum_jj\hat{n}_j+{i}V_1t\sum_{j\neq k}\frac{(-1)^{k+j}}{2}\hat{n}_k\hat{n}_j\right],
\end{equation}
with $\theta(t)=\int_0^t d\tau F(\tau)=F_1\sin\omega t+F_2\sin2\omega t+F_3\sin3\omega t$, such that the Hamiltonian in the rotating frame reads
\begin{equation}
\widetilde{H}{=}U(t)\hat{H}U^\dagger(t)- iU(t)\frac{\partial U^\dagger(t)}{\partial t}{=}\sum_{\langle jk\rangle}\hat{c}_j^\dagger\hat{B}_{jk}(t)\hat{c}_k.
\end{equation}
The global density dependence is contained in the operator
\begin{equation}
     \hat{B}_{jk}(t){=} J_{jk}\exp\left[i\theta(t)(j-k)+iV_1t\hat{\Delta}_{jk}/2\right].
\end{equation} 
We require that $V_1$ must be integer multiples of $\omega$, such that the system remains periodically modulated in the rotating frame.
The lowest-order contribution to the effective Hamiltonian follows as
\begin{equation}
\begin{aligned}
    \hat{H}_{\text{eff}}&{=}\frac{1}{T}\int_0^Tdt\widetilde{H}(t)\\
    &{=}\sum_{j}\hat{c}_{j+1}^\dagger\frac{1}{T}\int_0^TdtJ_{jk}\exp\left[i(F_1\sin\omega t+F_2\sin2\omega t+F_3\sin3\omega)+i\frac{\hat{\Delta}_{jk}}{2}\omega t\right]\hat{c}_j\\
    &{=}\sum_{j}\hat{c}_{j+1}^\dagger\mathcal{J}_{-\frac{V_1\hat{\Delta}_{jk}}{2\omega}}(F_1,F_2,F_3)\hat{c}_j,
\end{aligned}
\end{equation}
where the order of the Bessel function $-\frac{V_1\hat{\Delta}_{jk}}{2\omega}$ must be an integer according to the definition of the $N$-dimensional Bessel function. As $\hat{\Delta}_{jk}$ equals $2(-1)^{j}\sum_l(-1)^l\hat{n}_l$ and $\hat{n}_l$ is always 0 or 1 in our setup, $\Delta_{jk}$ must be an even number. Therefore, we choose $V_1=\omega$, leading to 
\begin{equation}
\hat{H}_{\text{eff}}{=}\sum_{j}\hat{c}_{j+1}^\dagger\mathcal{J}_{-\frac{\hat{\Delta}_{jk}}{2}}(F_1,F_2,F_3)\hat{c}_j.
\end{equation}
Having realized the global-density–dependent tunneling, a Toffoli gate can be constructed in a manner similar to that described in the main text. We again take the 4-qubit case as an example. After calculation, the driving parameters are required to satisfy
\begin{equation}
    {J}_{-3}(F_1,F_2,F_3)={J}_{-1}(F_1,F_2,F_3)={J}_{1}(F_1,F_2,F_3)=0.
\end{equation}
As demonstrated above, the variables of the three Bessel functions are all the same. Only the order of the Bessel functions varies. Solving these three equations amounts to searching for common zeros of three Bessel functions of distinct orders. However, we can still use a similar numerical optimization method as in our main text. Defining the cost function $g=|{J}_{-3}(F_1,F_2,F_3)|+|{J}_{-1}(F_1,F_2,F_3)|+|{J}_{1}(F_1,F_2,F_3)|$ and using the CG method with multiple starting points, a numerical solution can be found as:
\begin{equation}
    F_3=-5.97,F_2=-6.90,F_1=-6.62,g\approx10^{-6}.
\end{equation}

\section{Experimental feasibility}
As discussed in the main text, our protocol relies on a strongly driven system, where the amplitude of long-range interactions can scale up with the driving frequency. Here we discuss potential experimental parameters for the realization of the target global kinetic constraints.

For simplicity, we consider the same driving protocol that leads to Eq. (2) in the main text in the $N$-qubit system. Different hopping channels now read
\begin{equation*}
    \mathcal{J}(F_3,F_2,(N-1)V_1),\,\mathcal{J}(F_3,F_2,(N-3)V_1),...,\,\mathcal{J}(F_3,F_2,-(N-1)V_1).
\end{equation*}
Suppose that we want to suppress the hopping channel to impose the global kinetic constraints, 
\begin{equation*}
\label{eq.SMcondition}
    \mathcal{J}(F_3,F_2,(N-1)V_1)=0.
\end{equation*}
As shown in Fig.~2, the optimization results are $F_3=1,F_2=2, (N-1)V_1=4.26.$
More generally, this optimization corresponds to a root-finding problem and one can always find a root of this Bessel function by numerical optimization, such that $(N-1)V_1$ takes a value of order $\mathcal{O}(1)$. Therefore, one can estimate that the required value of $V_1$ scales as $V_1\sim O(\frac{1}{N})$. Crucially, this suggests that a weak driving strength of the global interaction already suffices to realize the desired global kinetic constraints, due to the fact that the effective interaction strength can be amplified by increasing the number of atoms in the system.

Now we provide more details regarding the driving parameters that are experimentally accessible. 
The natural energy scale of the system is set by the hopping amplitude $t$, which in experiments ranges from 10 Hz to 100 Hz. { Concretely, for a lattice depth of $10E_r$ the tunneling amplitude is $t_q=85$~Hz. According to Ref.~\cite{landig2016quantum}, at the cavity detuning $\Delta \approx -18$~MHz, the effective long-range interaction gets amplified by the atomic number $N$, leading to approximate interaction strength $2\pi \times 3.5$~kHz.
To realize our driving protocol, we can set the modulation frequency to $\omega = 850$~Hz, ten times the hopping unit, to realize a high-frequency drive. We further consider the particle number $N=1000$ and $V_1\approx 4.26\times10^{-3}$ such that the aforementioned optimal condition can be satisfied. 
We also note that the band gap of the optical lattice is around $24$~kHz, and hence the single-band approximation remains valid.
Therefore, our scheme lies within an experimentally feasible regime.

In practice, the system also suffers from particle loss, spatial disorder and other experimental imperfections. We leave a detailed stability analysis to future works.

\end{document}